\documentclass[english,keywords,amsmath,amssymb,twocolumn]{revtex4}
\usepackage[T1]{fontenc}
\usepackage[latin1]{inputenc}
\usepackage{braket}
\usepackage{babel}%
\usepackage{graphicx}
\usepackage{color}
\usepackage{bm}
\usepackage{longtable}
\usepackage{amsmath}
\usepackage{amsfonts}
\usepackage{dsfont}
\usepackage{amssymb}
\usepackage{hyperref}
\begin{document}
\title{Local $PT$-symmetric evolutions on separable states and violation of no-signaling}
\author{Asmita Kumari}
\author{Ujjwal Sen}
\affiliation{Harish-Chandra Research Institute, HBNI, Chhatnag Road, Jhunsi, Allahabad 211 019, India}

\begin{abstract}

We show that local \(PT\)-symmetric evolutions can lead to violation of the no-signaling principle for separable and even classically correlated bipartite shared quantum states. For classically correlated states, specially chosen \(PT\)-symmetric  operations from a set of zero volume can also preserve the principle.   The violations can be removed by using a \(CPT\) inner product instead of the traditional one. 


\end{abstract}

\maketitle
\section{Introduction}
 
It is usually assumed in  textbook quantum mechanics that the observables to be measured are hermitian operators. The possible outcomes of the measurement of these hermitian operators are then the eigenvectors - or clusters of them - and for systematic bookkeeping of what transpired in a measurement, we use the corresponding real eigenvalues. The eigenvectors associated with the eigenvalues  form a complete orthonormal set. The evolution corresponding to hermitian Hamiltonians are unitary~\cite{balo-kothai-tomar-desh}. However, non-hermitian 
observables have also been found to have useful applications, and
%
%
in recent years, such Hamiltonians with real eigenvalues have garnered 
considerable interest. ``\(PT\)-symmetric'' Hamiltonians~\cite{bender98} form an important example of such a class of non-hermitian Hamiltonians,
which respect  parity and time-reversal symmetry.
Although \(PT\)-symmetric Hamiltonians  have real eigenvalues, the eigenvectors corresponding to distinct eigenvalues may not be orthogonal, and  evolutions governed by \(PT\)-symmetric Hamiltonians are non-unitary. 
See \cite{curt07,brody14,rosa18,zel20,eijey, bender02,brody12, current-nei1, current-nei2} and references therein for further aspects.

 Although the interpretation of  
 \(PT\)-symmetric quantum mechanics remain contentious, it has been used in several areas~\cite{chang,graefe,kreibich,deffner, bender17,gunther,lee, bro16, japa, amar-balar-kichhu-chhilo-na,pati15,pati19,pati14,dey19,mitt,asmi}. Experiments related to such evolutions have been reported in~\cite{nagarik-klantite, ruter, preme-poRa-baran, 1, 2, 3, 4, 5, 6, 7, 8, 9, 10, 11, 12, 13, 14, 15, 16, 17, 18}.  Recently,  Lee \emph{et al.} in Ref.~\cite{lee} showed that the local operation of the evolution operator governed by a \(PT\)-symmetric Hamiltonian, on a bipartite entangled shared state~\cite{HHHH},  violates the no-signaling principle. This result 
 has been experimentally tested by Tang \emph{et al.}~\cite{nagarik-klantite}.
A post-selection process was used to simulate the \(PT\)-symmetric evolution, and without the post-selection, the violation of no-signaling does not appear.

The no-signaling principle requires that instantaneous communication of information at a distance is not allowed~\cite{diganta}. Local non-relativistic quantum evolution on shared quantum states do not allow the violation of this principle.  
The issue of violation of the no-signaling principle via local \(PT\)-symmetric evolutions has been re-addressed, and in particular, in Refs. \cite{bro16, japa,suna-koi-kahani}, it has been shown that utilizing a ``\( CPT\) inner product'', instead of the standard inner product of separable Hilbert spaces, can restore  the no-signaling principle within \(PT\)-symmetric quantum mechanics. See also~\cite{bro16,japa, bender02, current-nei1, current-nei2, lalon-ki-jaat, tumi-rabe-nirabe, ami-ki-niye-thaki, bender07} in this regard.



The violation of the no-signaling principle via local \(PT\)-symmetric evolutions has been dealt with by using shared quantum states that are entangled and pure. We show that even mixed states can be used to attain the same violation. Moreover, such mixed states can even be separable. We derive this by using the Werner states~\cite{swapna-madhur-mohe}, which is a single-parameter family of shared two-party states, and contains both entangled and separable states. 
We find that all nontrivial Werner states lead to the violation. Separable Werner states however contain a type of quantum correlation called quantum discord~\cite{kul-nai-kinar-nai}, which is somewhat subtler than entanglement. We however identify another class of shared two-party states that do not have any type of quantum correlation - but are classically correlated - that lead to signaling when \(PT\)-symmetric evolution is applied locally. 
Subsequently, we show that 
by using  the \(CPT\) inner product, one can reinstate the no-signaling principle in all cases considered.


The rest of the paper is arranged as follows. In Sec.~\ref{dui}, we discuss the most general two-dimensional $PT$-symmetric Hamiltonian. Violation of the no-signaling principle in \(PT\)-symmetric quantum mechanics for shared Werner states is considered in  
Sec.~\ref{tin}. The case of classically correlated states without shared quantum discord is taken up in Sec.~\ref{char}.
The \(CPT\) inner product consideration is presented in Sec.~\ref{pienc}.
We present a conclusion in Sec.~\ref{choo}.

\section{Hamiltonian of a $PT$-symmetric spin-1/2 system}
\label{dui}
 Let us begin by considering the  most general $PT$-symmetric Hamiltonian of a system described on a two-dimensional complex Hilbert space. Let it be denoted by 
 \(J H_{PT}\), where \(H_{PT}\) given by \cite{eijey}
\begin{eqnarray}
\label{evo} 
H_{PT}=\left(
\begin{array}{cc}
  r+t \cos\xi-i s \sin\xi & i s \cos\xi+t \sin\xi \\
 i s \cos\xi+t \sin\xi & r-t \cos\xi+i s \sin\xi \\
\end{array}
\right).\nonumber\\
\end{eqnarray}
The parameters $r, s, t, \xi$ containing in the Hamiltonian are real. The constant, \(J\), is a real number that is nonzero and that has the unit of ``energy''. This implies that the other parameters are dimension-free. The 
parity operator, \(P\), is given by 
\begin{eqnarray}
\label{phaTkabajir-deshe} 
P=\left(
\begin{array}{cc}
  \cos\tilde{\phi} & \sin\tilde{\phi} \\
\sin\tilde{\phi} & -\cos\tilde{\phi} \\
\end{array}
\right), 
\end{eqnarray}
where \(\tilde{\phi}\) is real. 
The time-reversal operator, \(T\), is complex conjugation in the computational basis, with the latter being the eigenbasis of the Pauli \(\sigma_z\) operator~\cite{eijey}. 
We are interested in those $H_{PT}$ for which the eigenvalues are real, so that we require \(s^2 \leq t^2\). Setting $\sin \alpha = s/t$, we find that 
the 
eigenvalues of $J H_{PT}$ are $E_{\pm} = J(r \pm t \cos\alpha)$. The  respective eigenstates are given by 
eigenstates being 
\begin{eqnarray} 
|E_{\pm} (\alpha) \rangle = \frac{1}{N_1}\left(
\begin{array}{c}
 \pm \sin{\xi}\sin{\alpha}- i \sin{\alpha}  \\
 1 \mp \cos\alpha \cos{\xi} \\
\end{array}
\right),   
\end{eqnarray}
where $N_1=\sqrt{2 (1\mp \cos\alpha)\cos\alpha}$. 
Unlike the eigenstates of a hermitian Hamiltonian, the eigenstates of \(H_{PT}\) are nonorthogonal. 
Also, the eigenvectors fall on each other 
at the ``branch points'', $\alpha = \pm \pi /2$.
We will consider \(PT\)-symmetric Hamiltonians away from their branch points. 
%
%
See Refs.~\cite{lalon-ki-jaat, koran-bible-upanishad}.

\section{Werner states and violation of no-signaling}
\label{tin}

Lee and co-authors in Ref. \cite{lee} have shown that local evolutions governed by  \(PT\)-symmetric Hamiltonians on bipartite entangled states
can result in the violation of the no-signaling principle. In this section, we do the same analysis with the local evolution corresponding to the most general \(PT\)-symmetric Hamiltonian acting on an arbitrary (but fixed) shared Werner state.

 Let Alice and Bob be in two space-like separated locations, sharing the Werner state given by
\begin{eqnarray}
\rho_{AB_w}= p |\psi \rangle \langle \psi|_{AB} + \frac{1-p}{4}\mathbb{I}_2 \otimes \mathbb{I}_2,
\end{eqnarray}
where $|\psi \rangle_{AB} = \frac{1}{\sqrt{2}}(|00 \rangle + |11 \rangle)$, and \(|0\rangle\) and \(|1\rangle\) are eigenstates of the Pauli \(\sigma_z\) operator.
\(\mathbb{I}_2\) is the identity operator on the qubit Hilbert space. For \(\rho_{AB_w}\) to be a quantum state, we require \(p\) to lie in the range \([-1/3,1]\).
The no-signaling principle demands that the probability distribution of outcomes of an arbitrary measurement at  Bob's end is unaffected by Alice's choice of operations. In other words, the no-signaling principle is satisfied if the reduced density matrix of Bob before as well as after the local operation of Alice are the same.  Before a local operation, the state at Bob's end is given by
\begin{eqnarray}
 \rho_{B_w} =\frac{1}{2}\left(
\begin{array}{cc}
  1 & 0 \\
 0 & 1 \\
\end{array} 
\right)= \frac{\mathbb{I}_2}{2}. 
\end{eqnarray}
Now, let us assume that Alice evolves her part of the shared Werner state by using the $PT$-symmetric Hamiltonian given in Eq.~(\ref{evo}). 
If the non-unitary evolution by Alice is $U_A =e^{-i H_{PT} \tau}$, where \(\tau = J \tau{'}/\hbar\), with \(\tau{'}\) playing the role of ``time'', then the composite density matrix reduces to
\begin{eqnarray}
\rho_{AB_w}^{U_A} = U_A \otimes \mathbb{I}_2 \rho_{AB_w} U_A^{\dagger} \otimes  \mathbb{I}_2.
\end{eqnarray}
In order to check the validity of no-signaling principle, we take the partial trace over Alice's part. Then the reduced state of Bob, after renormalization, is obtained as
\begin{eqnarray}
\rho_{B_w}^{U_A} &=& \mbox{Tr}_A[\rho_{AB_w}^{U_A}] = \frac{1}{N_2} \left(
\begin{array}{cc}
  w_{11} & w_{12} \\
 w_{21} & w_{22} \\
\end{array}
\right), 
\end{eqnarray}
where $N_2 = 2 \sec ^2\alpha  \sin ^2t_1+\cos (2 t_1)$, and
\begin{eqnarray}
\label{jene-shune-bish}
\nonumber
w_{11} &=&  \frac{1}{2}\left[
N_2 
- p \tan \alpha  \sin (2 t_1) \sin \xi \right], \\ \nonumber
w_{22} &=& 
\frac{1}{2}\left[ 
N_2
+p \tan \alpha  \sin (2 t_1) \sin \xi\right], \\ \nonumber
w_{21} &=& 
2 p \tan \alpha  \sin t_1 \left(\cos t_1 \cos \xi 
+ i \sec \alpha  \sin t_1 \right), \\ 
w_{12} &=& 
w_{21}^*.
\end{eqnarray}
Here $t_1 = t \tau \cos\alpha$. To satisfy the no-signaling principle, we require
$\rho_{B_w}^{U_A} = \mathbb{I}_{2}/2$. 
We exclude the branch points at 
\(\alpha = \pm \pi/2\).
By comparing the elements of  $\rho_{B_w}^{U_A} $ with those of \(\rho_{B_w}\), we can see that no-signaling principle is satisfied only if \(\alpha =0\) or \(t_1 =0\) or \(p=0\). For \(\alpha =0\), the Hamiltonian \(H_{PT}\) becomes hermitian. And \(t_1 = 0\) gives \(t=0\), which in turn  leads to a Hamiltonian, \(H_{PT}\), that is hermitian,  as then \(s=0\), or \(\tau =0\), which implies that the evolution did not occur.  We therefore find that  all \(PT\)-symmetric non-hermitian Hamiltonians with real eigenvalues lead to signaling of information by using the pre-shared Werner state (unless \(p=0\)), just as for the maximally entangled state shown in \cite{lee}. A maximally entangled state is obtained from the family of Werner states by setting \(p=1\).

It is known that the family of Werner states contains entangled as well as separable states. Indeed, it is entangled in the range \(p \in (1/3,1]\), while is separable for \(p\in[-1/3,1/3]\)~\cite{swapna-madhur-mohe}. Therefore, the violation of no-signaling can be afforded by shared states without any entanglement, for local \(PT\)-symmetric evolutions.


\section{Classically correlated states and violation of no-signaling}
\label{char}
In the preceding section, we have seen that separable states can also offer violation of the no-signaling principle, for local \(PT\)-symmetric evolutions. We will see in this section that even weaker correlations suffice for the same violation to occur. To deal with task, we will need some concepts in quantum correlations, and we digress in the next paragraph for a discussion about the same.

A bipartite quantum state, \(\rho_{AB}\), shared between two observers, \(A\) and \(B\), is said to be entangled if it cannot be expressed in separable form, viz. 
\begin{equation}
\rho_{AB} = \sum_i p_i \tilde{\rho}^i_A \otimes \tilde{\tilde{\rho}}^i_B,
\end{equation}
where \(\{p_i\}\) form a probability distribution, and \(\tilde{\rho}^i_A\) (\(\tilde{\tilde{\rho}}^i_B\)) are density matrices of the physical system in possession of \(A\) (\(B\))~\cite{HHHH}. It has however been realized that quantum correlations in shared systems can also be conceptualized independent of the entanglement-separability paradigm~\cite{kul-nai-kinar-nai}, with one of the popular ones being quantum discord. It is defined as the difference between quantized versions of two classically equivalent versions of mutual information of measured bipartite quantum states. It was found that while quantum discord is identical with entanglement for pure bipartite states, there exists separable mixed bipartite states with a nonzero quantum discord. 
States with zero quantum discord have been referred to as ``classically correlated states'', and forms a (non-convex and zero-volume) set, strictly within the set of separable states.

In particular, all Werner states for \(p\ne0\) can be shown to have a nonzero  quantum discord. However, we show in this section that bipartite quantum states having a vanishing quantum discord can also lead to violation of no-signaling when acted locally with \(PT\)-symmetric evolutions.

Let us therefore assume again that Alice and Bob are at two space-like separated locations, 
sharing the classically correlated state given by
\begin{eqnarray}
\label{sep}
\sigma_{AB_s}= p |00 \rangle \langle 00|_{AB} + (1-p) |11 \rangle \langle 11|_{AB}.
\end{eqnarray}
where we must now require $0 \leq p \leq 1$. For the case of the state given in Eq.~(\ref{sep}), the no-signaling principle is satisfied if the reduced density matrix of Bob before as well as after a local \(PT\)-symmetric operation of Alice, are the same. Before the operation, Bob's state is
\begin{eqnarray}
\label{nss10}
 \sigma_{B_s} =\left(
\begin{array}{cc}
  p & 0 \\
 0 & 1-p \\
\end{array} 
\right). 
\end{eqnarray}
If Alice evolves her part of the classically correlated state by  using the $PT$-symmetric Hamiltonian given in Eq.~(\ref{evo}), then the composite system evolves to 
\begin{eqnarray}
 \nonumber
\sigma^{U_A}_{AB_s}= p U|0 \rangle \langle 0|U^{\dagger} \otimes |0 \rangle \langle 0| + (1-p) U|1 \rangle \langle 1|U^{\dagger} \otimes |1 \rangle \langle 1|.
\\
\end{eqnarray} 
Our next task is to trace out Alice's part from this evolved state, which results in the post-operation reduced density matrix of Bob given by 
\begin{eqnarray}
\label{nss11}
 \sigma^{U_A}_{B_s} =\frac{1}{N_3}\left(
\begin{array}{cc}
  p\langle 0|U^{\dagger}U|0 \rangle & 0 \\
 0 & (1-p ) \langle 1|U^{\dagger}U|1 \rangle \\
\end{array} 
\right), 
\end{eqnarray}
where $N_3 = (1-2 p) \tan \alpha  \sin \xi  \sin (2 t_1)+2 \sec^2 \alpha  \sin^2 t_1+\cos (2 t_1) $.
The nonzero elements of the density matrix $ \sigma^{U_A}_{B}$ require the following expressions:
\begin{eqnarray}
\nonumber
 \langle 0|U^{\dagger}U|0 \rangle &=& 2 \sec ^2\alpha  \sin ^2 t_1+\cos (2 t_1)\\ \nonumber &&-  \tan \alpha  \sin (2 t_1) \sin \xi,  \\ \nonumber
 \langle 1|U^{\dagger}U|1 \rangle &=& 2 \sec ^2 \alpha  \sin ^2 t_1 +\cos (2 t_1) \\ \nonumber &&+ \tan \alpha  \sin (2 t_1) \sin \xi.
  \end{eqnarray}
Here, it should be noted that $U^{\dagger}U \neq \mathbb{I}_2$. Using the expressions for $\langle 0|U^{\dagger}U|0 \rangle $ and $\langle 1|U^{\dagger}U|1 \rangle$, and comparing the pre- and post-operation density matrices at Bob's end, we find that the no-signaling requirement is satisfied if $\xi=0$. 
However, outside the surface given by  $\xi = 0$,
no-signaling is satisfied only if \(\alpha =0\) or \(t_1 =0\) or \(p=0,1\). As mentioned before,  \(\alpha =0\) leads to a hermitian \(H_{PT}\). \(t_1 = 0\) implies \(t=0\), which again leads to a hermitian \(H_{PT}\) as then \(s=0\), or \(\tau =0\), which means that the evolution did not happen.

We therefore find that even classically correlated states can be used to obtain violation of no-signaling for local \(PT\)-symmetric evolutions. There is however a distinct difference with the considerations for the Werner class. For Werner states, all \(PT\)-symmetric evolutions lead to violation of no-signaling, while for the class of states discussed here, there is a (zero-volume) surface in the parameter space of \(PT\)-symmetric evolutions on which the no-signaling principle is satisfied. 



\section{\(CPT\) inner product and no-signaling principle}
\label{pienc}
It has previously been shown in the literature that using a \(CPT\) inner product can remove the violation of no-signaling in the case of shared entangled states and local \(PT\)-symmetric evolutions. In this section,  we  demonstrate that violation of the no-signaling principle can be avoided both for the Werner state and for the classically correlated  state using a \(CPT \) inner product. It is known that evolution generated by a non-hermitian $PT$-symmetric Hamiltonian is not unitary. This is the basic reason for  the violation of the no-signaling principle. However, if a \(CPT \) inner product is used appropriately, the time evolution corresponding to a $PT$-symmetric Hamiltonian becomes unitary. In order to do so, let us introduce a charge conjugation operator, $C$, defined as
\begin{eqnarray}
\label{c}
&&C= |E_{+} (\alpha) \rangle \langle E_{+} (\alpha)| + |E_{-} (\alpha) \rangle \langle E_{-} (\alpha)| \nonumber \\ \nonumber
&&=\frac{1}{N_4}\left(
\begin{array}{cc}
 \cos \xi -i \sin \alpha  \sin \xi  & i \cos \xi  \sin \alpha +\sin \xi  \\
 i \cos \xi  \sin \alpha +\sin \xi  & i \sin \alpha  \sin \xi -\cos \xi  \\
\end{array}
\right),\\
\end{eqnarray}
where $N_4 = \cos \alpha $. The eigenstates of $H_{PT}$ are simultaneous eigenstates of the $C$ operator. Indeed, the matrix $C$ commutes with  $H_{PT}$ and $PT$, and $C^2 =1$. Using the matrix representation of the parity operation defined in Eq.~(\ref{phaTkabajir-deshe}) and of the charge conjugation operator, $C$, in Eq.~(\ref{c}), we obtain
\begin{eqnarray}
\label{cp}
CP = \left(
\begin{array}{cc}
 \sec \alpha  & -i \tan \alpha  \\
 i \tan \alpha  & \sec \alpha  \\
\end{array}
\right).
\end{eqnarray}
We now define the time evolution operator 
for the $PT$-symmetric Hamiltonian, \(H_{PT}\), as
\begin{eqnarray}
\mathcal{U} = e^{-Q/2} e^{-i H_{PT} \tau}  e^{Q/2},
\end{eqnarray}
where $ e^{Q} = C P $. Using the representation of $H_{PT}$ in Eq.~(\ref{evo}) and that 
of  $CP$ given in Eq.~(\ref{cp}), the evolution operator, $\mathcal{U}$, 
reduces to 
\begin{eqnarray}
&&\mathcal{U} =\nonumber \\ \nonumber && e^{-i r \tau } \left(
\begin{array}{cc}
 \cos t_1-i \cos \xi  \sin t_1 & -i \sin \xi  \sin t_1 \\
 -i \sin \xi  \sin t_1 & \cos t_1+i \cos \xi  \sin t_1 \\
\end{array}
\right).\\
 \end{eqnarray}
It can easily be checked that 
$\mathcal{U}$ is an unitary operator.
See e.g. Refs.~\cite{bender07, kaljayee} for further details.
Now, if Alice uses $\mathcal{U}$ for the evolution of her local state, then for the case of the shared Werner state, we obtain
\begin{eqnarray}
\label{ns1}
 \rho_{B_w} = \rho_{B_w}^{\mathcal{U}} =\frac{1}{2}\left(
\begin{array}{cc}
  1 & 0 \\
 0 & 1 \\
\end{array} 
\right),
\end{eqnarray}
 and hence, the no-signaling principle is satisfied. Similarly, for the case of the classically correlated state, if Alice evolves her local state using $\mathcal{U}$, then the state of Bob before and after the measurement are equal:
\begin{eqnarray}
\label{nss12}
 \sigma_{B_s}= \sigma^{\mathcal{U}}_{B_s} =\left(
\begin{array}{cc}
  p & 0 \\
 0 & 1-p \\
\end{array} 
\right),
\end{eqnarray}
and hence again no-signaling is retrieved. 
Indeed, for any unitary \(\tilde{U}_A\) on Alice's Hilbert space and for any shared quantum state \(\varrho_{AB}\) between Alice and Bob, \(\mbox{tr}_A \left(\tilde{U}_A \otimes \mathbb{I}_B \varrho_{AB}\tilde{U}^\dagger_A \otimes \mathbb{I}_B \right)= \mbox{tr}_A \left(\varrho_{AB}\right)\), where \(\mathbb{I}_B\) is the identity operator at Bob's end.

\section{Conclusion}
\label{choo}
$PT$-symmetric quantum mechanics
remains a contentious theory and exhibits a host of 
strange behavior. Violation of the no-signaling principle is one of them. In \cite{lee}, it was shown that a local $PT$-symmetric operation on a maximally entangled state violates the no-signaling principle. We found that the entanglement content of the shared state is independent of the no-signaling violation, and the same can also be obtained for states with zero entanglement. We went on to show that even classically correlated states that does not possess any quantum discord, a quantum correlation concept that is independent of entanglement, can  offer violation of the no-signaling principle for local \(PT\)-symmetric evolutions. 
We subsequently found that using a \(CPT\) inner product can reinstate the no-signaling principle, just like it was found to do so for entangled states. 


\acknowledgements
 We acknowledge partial support from the Department of Science and Technology, Government of India through the QuEST grant (grant number DST/ICPS/QUST/Theme-3/2019/120).

\end{document}